\documentstyle[12pt,titlepage,epsf]{article}
\def\_{\rule{.3em}{.15ex}} 
\def\bbuildrel#1_#2^#3{\mathrel{\mathop{\kern 0pt#1}\limits_{#2}^{#3}}}
\newcommand{\scs}{\scriptscriptstyle}
\newcommand{\be}{\begin{equation}}
\newcommand{\ee}{\end{equation}}
\newcommand{\bea}{\begin{eqnarray}}
\newcommand{\eea}{\end{eqnarray}}
\newcommand{\f}{\frac}
\newcommand{\me}[1]{\langle#1\rangle}
\newcommand{\al}{\alpha_s}
\newcommand{\e}{\epsilon}

\newcommand{\s}{\hspace{1cm}}
\newcommand{\p}{\hspace{0.5cm}}

\begin{document}
\begin{titlepage}

 \begin{flushright}
  {\bf MPI/PhT/97-51\\
       TTP97-44$^{^{\dagger}}$ \\
       ZU-TH-17/97\\
       TUM-HEP-285/97\\       
       IFT-12/97\\
      hep-ph/9711280\\
}
 \end{flushright}

 \begin{center}
  \vspace{0.6in}

\setlength {\baselineskip}{0.3in}
  {\bf \Large 
$|\Delta F| = 1$ Nonleptonic Effective Hamiltonian in a Simpler Scheme}
\vspace{2cm} \\
\setlength {\baselineskip}{0.2in}

{\large  Konstantin Chetyrkin$^{^{1,\star}}$, 
         Miko{\l}aj Misiak$^{^{2}}$
         and Manfred M{\"u}nz$^{^{3}}$}\\

\vspace{0.2in}
$^{^{1}}${\it Institut f{\"u}r Theoretische Teilchenphysik, Universit{\"a}t Karlsruhe, \\
                        D-76128 Karlsruhe, Germany}

\vspace{0.2in}
$^{^{2}}${\it Institute of Theoretical Physics, Warsaw University,\\
		 PL-00-681 Warsaw, Poland}

\vspace{0.2in}
$^{^{3}}${\it Physik Department, Technische Universit{\"a}t M{\"u}nchen,\\
                         D-85748 Garching, Germany}

\vspace{2cm} 
{\bf Abstract \\} 
\end{center} 
\setlength{\baselineskip}{0.3in} 

	We consider $|\Delta F| = 1 \;\;\;(F = S,C$ or $B)\;\;$
nonleptonic effective hamiltonian in a renormalization scheme which
allows to consistently use fully anticommuting $\gamma_5$ at any
number of loops, but at the leading order in the Fermi coupling $G_F$.
We calculate two-loop anomalous dimensions and one-loop matching
conditions for the effective operators in this scheme. Finally, we
transform our results to one of the previously used renormalization
schemes, and find agreement with the original calculations.

\vspace{1cm}

\setlength {\baselineskip}{0.2in}
\noindent \underline{\hspace{2in}}\\ 
\noindent 
$^{^{\dagger}}${\footnotesize The
  complete postscript file of this
  preprint, including figures, is available via anonymous ftp at
  www-ttp.physik.uni-karlsruhe.de (129.13.102.139) as /ttp97-44/ttp97-44.ps
  or via www at http://www-ttp.physik.uni-karlsruhe.de/cgi-bin/preprints.}

\noindent
$^\star$ {\footnotesize 
Permanent address: Institute of Nuclear Research,
Russian Academy of Sciences, Moscow 117312, Russia.}

\end{titlepage} 

\setlength{\baselineskip}{0.3in}

\noindent {\bf 1. Introduction}

	Performing multiloop calculations is usually most
convenient within the framework of dimensional regularization.
However, when chiral fermions are considered, one often
encounters technical difficulties with extending $\gamma_5$ to
$D = 4 - 2 \e$ dimensions. These difficulties are related
to Dirac traces like 
\be \label{nasty.trace}
T_{\mu\nu\rho\sigma} = Tr(\gamma_{\alpha} 
\gamma_{\mu} \gamma_{\nu} \gamma_{\rho} \gamma_{\sigma}
   \gamma^{\alpha} \gamma_5).
\ee
Such a trace is not uniquely defined in $D$ dimensions, so long
as one assumes that $\gamma_5$ anticommutes with all the other
gamma matrices. One can easily check that the totally
antisymmetric part of  $T_{\mu\nu\rho\sigma}$ equals to
\be \label{antisym.trace}
T_{[\mu\nu\rho\sigma]} = -D \; Tr( 
\gamma_{[\mu} \gamma_{\nu} \gamma_{\rho} \gamma_{\sigma]} \gamma_5)
\ee
when cyclicity of the trace and $\{\gamma_{\alpha},\gamma_5\} =
0$ is used, or to
\be
T_{[\mu\nu\rho\sigma]} = (D-8) \; Tr(
\gamma_{[\mu} \gamma_{\nu} \gamma_{\rho} \gamma_{\sigma]} \gamma_5)
\ee
when the two contracted $\gamma_{\alpha}$'s are brought to each other
through the whole string of other gamma matrices with use
of $\{\gamma_{\alpha},\gamma_{\beta}\} = 2 g_{\alpha\beta}$.

	These two results agree with each other only when $D=4$ or
when the whole trace vanishes. Neither ambiguous results for $D \neq
4$ nor the vanishing trace can be accepted in the Dirac algebra in a
consistent dimensional regularization scheme. The Dirac algebra in an
acceptable scheme must satisfy the following requirements:
\begin{itemize}
\item{(i)} It should give unique results independently of whether some
diagram is considered as a subdiagram, and independently of the order
in which subdiagrams are calculated. Otherwise subdivergences could
not be properly subtracted in multiloop diagrams.
\item{(ii)} When $D \to 4$, its rules must analytically tend to the
rules of 4-dimensional Dirac algebra.
\end{itemize}

	Several schemes satisfying these requirements have been
proposed \cite{HV72}--\cite{KKS92}. What they have in common is
that their use requires complicated or tedious algebraic
manipulations.

	Calculations become much simpler if occurrence of traces
with $\gamma_5$ can be avoided in a given problem. Then, one can
define the $D$-dimensional $\gamma_5$ as 
\be \label{nice.gamma5}
\gamma_5 = i^{\f{(D-1)(D-2)}{2}} \gamma^0 ... \gamma^{{\scs D}-1}.
\ee
With such an explicit definition at hand, the requirement (i) is
automatically satisfied. The requirement (ii) would not be satisfied
for diagrams containing traces with $\gamma_5$. (The trace
(\ref{antisym.trace}) would then identically vanish for
$D\neq4$). However, in the absence of traces with $\gamma_5$, the
requirement (ii) is satisfied, too. The matrix defined in
eqn.~(\ref{nice.gamma5}) anticommutes with all the remaining gamma
matrices, and its square is equal to unity.

	Flavor changing $|\Delta F| =1 \;\;\;(F = S,C$ or $B)\;\;$
processes in the Standard Model (SM) are described by Feynman diagrams
with no traces containing $\gamma_5$. This statement is true only at
the leading order in weak interactions, but to all orders in QCD and
QED.  While higher orders in weak interactions are usually negligible,
one is often interested in resumming large logarithms 
$\ln(M_W^2/m_{light}^2)$ from all orders of the QCD perturbation
series. This is most conveniently done with use of an effective
hamiltonian built out of light fields only. It is possible to define
this hamiltonian in such a way that no traces with $\gamma_5$ are 
introduced.

	In the standard approach to $|\Delta F| =1$ processes
\cite{GW79}, the applied form of the effective hamiltonian induced
appearance of traces with $\gamma_5$. It was harmless at one loop, but
caused many difficulties in the next-to-leading order (NLO) 2-loop
calculations \cite{BJLW92,BJLW93,CFMR94,CR93,CCRV94}. Applying the same
scheme in even higher order computations would be very
inefficient. Instead, it is better to start from the outset with a
modified form of the effective hamiltonian which is free of the
$\gamma_5$ problem.\footnote{
A similar issue concerning renormalization of four-quark operators
has been discussed in ref.~\cite{AC94}, in the context of QCD sum rules.}

	In the present paper, we present our calculation of the
next-to-leading QCD corrections to the $|\Delta F| = 1$ nonleptonic
effective hamiltonian. We use an operator basis in which traces with
$\gamma_5$ do not occur. Consequently, we are allowed to use fully
anticommuting $\gamma_5$ to all orders in QCD. Such an approach is
much more natural and useful than using the ``standard'' operator
basis \cite{BJLW92,BJLW93,CFMR94}. In our case, the two-loop
calculation can be made completely automatic and almost trivial. We do
not face subtleties related to the use of Fierz symmetry arguments in
$D$ dimensions, as it was the case when fully anticommuting $\gamma_5$
was used in the ``standard'' operator basis. The calculation is much
simpler than in the 't~Hooft-Veltman scheme \cite{HV72} for
$\gamma_5$, too. In this latter scheme, $D$-dimensional Lorentz
invariance is violated, which causes appearance of a large number of
complicated evanescent operators.

	Our main motivation for undertaking the present work was that
its outcome was directly applicable in our 3-loop calculation of NLO
corrections to $b \to X_s \gamma$ decay \cite{CMM97.1,CMM97.4}.
However, we find it useful to discuss the nonleptonic hamiltonian
separately from 3-loop $b \to X_s \gamma$ technicalities.

	Our calculation proceeds along the standard lines
\cite{BJLW92,BJLW93,CFMR94}. We evaluate tree-level and one-loop matching
conditions for the operators in the effective hamiltonian, as well as
their one- and two-loop anomalous dimensions. We only use another
basis of effective operators.  Even with our results at hand, it is
quite nontrivial to verify whether they agree with the previous
calculations \cite{BJLW92,BJLW93,CFMR94,ACMP81}. Such a comparison is
essential because of the phenomenological relevance of the considered
quantities.  Therefore, we devote a sizable part of the present article
to show how this verification is performed.

	Our paper is organized as follows. In the next section,
the effective hamiltonian is introduced. In section 3, we
calculate its Wilson coefficients and the necessary anomalous
dimension matrix. Section 4 is devoted to showing how our
results can be transformed to the previously used
renormalization scheme, and to verifying that we can indeed
confirm results of the previous calculations. The appendix
contains a list of the nonphysical counterterms relevant in
our calculation.

\ \\
\noindent {\bf 2. The effective hamiltonian}

	For definiteness, we shall consider here the $\Delta B =
-\Delta S = 1$ nonleptonic effective hamiltonian,\footnote{
The hamiltonian is assumed to conserve flavors other than $B$ and $S$.}
neglecting the small CKM matrix element $V_{ub}$. Generalization
to other $|\Delta F|~=~1$ processes and/or to processes where
$V_{ub}$ cannot be neglected is straightforward.

	The effective hamiltonian is built out of operators $P_i$
multiplied by their Wilson coefficients~$C_i$
\be \label{heff}
{\cal H}_{eff} = -\f{4 G_F}{\sqrt{2}} V^*_{ts} V_{tb} 
\sum_{i} C_i \left[P_i \;\; + \;\; (\mbox{counterterms})_i \right],
\ee
where $G_F$ is the Fermi coupling.

	The specific structure of the operators $P_i$ is determined
from the requirement that the hamiltonian reproduces the Standard
Model $\Delta B = -\Delta S = 1$ nonleptonic amplitudes at the leading
order in (external momenta)/$M_W$ and in the electroweak gauge
couplings, but to all orders in strong interactions. If this
requirement was applied to off-shell amplitudes, then EOM-vanishing
operators, i.e. operators which vanish by the QCD equations of motion
would need to be present among $P_i$. Here, we do not include such
operators. Consequently, our ${\cal H}_{eff}$ reproduces only on-shell
SM amplitudes (both partonic and hadronic ones) \cite{P80}. However,
we still require renormalizability of 1PI off-shell Green's functions
generated by ${\cal H}_{eff}$. Therefore, some EOM-vanishing operators
are present among counterterms.

	The above requirements imply that the set of operators
$\{P_i\}$ must be closed under QCD renormalization, up to
counterterms proportional to nonphysical operators, i.e.
operators whose renormalized matrix elements between physical
states vanish. Such nonphysical operators can be either
EOM-vanishing or evanescent (i.e. algebraically vanishing in
four dimensions) \cite{C84,evan}. Potentially, one could
also encounter BRS-exact operators (i.e. BRS-variations of some
other operators) as nonphysical counterterms \cite{C84,S94}.
However, they turn out to be unnecessary up to three loops for
the operators $P_i$ considered below.

	A set of operators $P_i$ which satisfies the imposed
requirements consists of dimension-six operators $P_1, ..., P_6$
\be \label{ope}
\begin{array}{rl}
P_1 = & (\bar{s}_L \gamma_{\mu} T^a c_L) (\bar{c}_L \gamma^{\mu} T^a b_L),
\vspace{0.2cm} \\
P_2 = & (\bar{s}_L \gamma_{\mu}     c_L) (\bar{c}_L \gamma^{\mu}     b_L),
\vspace{0.2cm} \\
P_3 = & (\bar{s}_L \gamma_{\mu}     b_L) \sum_q (\bar{q}\gamma^{\mu}     q),     
\vspace{0.2cm} \\
P_4 = & (\bar{s}_L \gamma_{\mu} T^a b_L) \sum_q (\bar{q}\gamma^{\mu} T^a q),    
\vspace{0.2cm} \\
P_5 = & (\bar{s}_L \gamma_{\mu_1}
                   \gamma_{\mu_2}
                   \gamma_{\mu_3}    b_L)\sum_q (\bar{q} \gamma^{\mu_1} 
                                                         \gamma^{\mu_2}
                                                         \gamma^{\mu_3}     q),     
\vspace{0.2cm} \\
P_6 = & (\bar{s}_L \gamma_{\mu_1}
                   \gamma_{\mu_2}
                   \gamma_{\mu_3} T^a b_L)\sum_q (\bar{q} \gamma^{\mu_1} 
                                                          \gamma^{\mu_2}
                                                          \gamma^{\mu_3} T^a q),
\end{array}
\ee
and one dimension-five operator proportional to\footnote{
The $s$-quark mass is neglected here. Therefore, $s_R$ decouples from 
flavor-changing processes.}     
\be \label{magn.mom}
(\bar{s}_L \sigma^{\mu\nu} T^a b_R) G^a_{\mu\nu},
\ee
called the chromomagnetic moment operator. Here, $T^a$ is the
$SU(3)_{color}$ generator, and $G^{a}_{\mu\nu}$ stands for the
gluon stress-energy tensor.

	Certainly, the above choice of the operator basis is not
unique. One could choose other linear combinations of the operators
and/or redefine them by adding some nonphysical ones.

\begin{figure}[ht] 
\centerline{
\epsfysize = 3cm
\epsffile{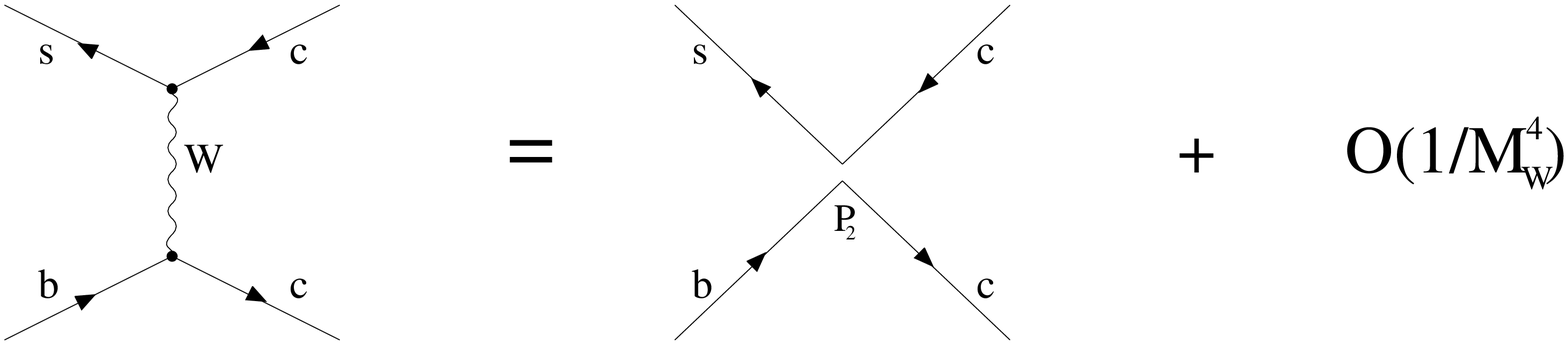}}
\caption{ Tree-level matching condition. \label{tree}}
\end{figure}

	Let us briefly argue why all the operators given in
eqns.~(\ref{ope}) and (\ref{magn.mom}) must be present in ${\cal
H}_{eff}$. The operator $P_2$ must be present because it enters the
matching condition shown in fig.~\ref{tree}. In this figure, the
$W$-boson exchange is approximated by the effective Fermi weak
interaction of quarks.

\begin{figure}[ht] 
\centerline{
\epsfysize = 2.5cm
\epsffile{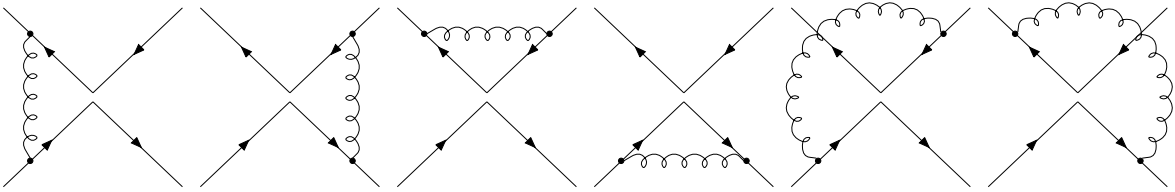}}
\caption{One-loop current-current diagrams. \label{1loop.current}}
\end{figure}

\begin{figure}[ht] 
\centerline{
\epsfysize = 3cm
\epsffile{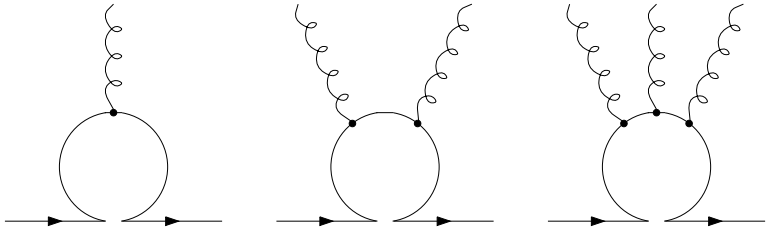}}
\caption{One-loop penguin diagrams with no Dirac trace. \label{1loop.side.penguin}}
\end{figure}

	All the other operators in eqns.~(\ref{ope}) and
(\ref{magn.mom}) are generated from $P_2$ in the process of QCD
renormalization of various amplitudes: The diagrams with
$P_2$-vertices shown in fig.~\ref{1loop.current} require
counterterms proportional to $P_1$, $P_2$ and to the nonphysical
evanescent operator $E^{(1)}_1$. (The relevant nonphysical
operators are collected together in the appendix.) All the
remaining divergent 1-loop 1PI diagrams with $P_2$-insertion are
shown in fig.~\ref{1loop.side.penguin}. They are renormalized by
a single counterterm proportional to 
\be \label{PDG}
(\bar{s}_L T^a \gamma_{\mu} b_L ) D_{\nu} G^{a\;\mu\nu}
\ee
where $D_{\nu}$ is the QCD covariant derivative. For convenience in
the future two-loop calculation, we write the above operator as a
linear combination of $P_4$ and the EOM-vanishing nonphysical operator
$N_1$ (see the appendix)
\be
N_1 = \f{1}{g} (\bar{s}_L T^a \gamma_{\mu} b_L ) D_{\nu} G^{a\;\mu\nu} 
\;\; + \;\; P_4.
\ee

	Inserting $P_4$ into the diagrams of fig.~\ref{1loop.current},
one finds that counterterms proportional to $P_5$ and $P_6$ are
necessary. Finally, the same diagrams with insertions of $P_6$ require
a counterterm proportional to $P_3$.

	The set of operators $\{P_1,...,P_6\}$ is closed under
one-loop QCD renormalization, up to nonphysical counterterms. A
set which closes to all orders in QCD is obtained by including
only one more operator, i.e. the chromomagnetic moment operator 
(\ref{magn.mom}). The fact that no more physical operators
of dimension 5 and 6 need to be present in ${\cal H}_{eff}$ can
be shown analogously to ref.~\cite{GSW90} by writing all the
possible $\Delta B =-\Delta S = 1$ operators allowed by gauge
symmetry and reducing them by the equations of motion.

	In the standard approach \cite{GW79,BJLW92,BJLW93,CFMR94}, triple
products of gamma matrices occuring in $P_5$ and $P_6$ have been
reduced with help of the four-dimensional identity
\be
\gamma_{\mu} \gamma_{\nu} \gamma_{\rho} 
\bbuildrel{=\!=\!=\!=}_{{\scs D} = 4}^{}
g_{\mu\nu} \gamma_{\rho} + g_{\nu\rho} \gamma_{\mu} - g_{\mu\rho} \gamma_{\nu}
+ i \epsilon_{\mu\nu\rho\sigma} \gamma^{\sigma} \gamma_5.
\ee
Doing this step in a two-loop calculation requires introducing several
more evanescent operators and leads to problematic traces with
$\gamma_5$. Therefore, we do not perform this step, and just leave the
operators $P_5$ and $P_6$ as they stand.\footnote{
An essentially identical procedure was used in the calculation of
one-loop corrections to the coefficient functions of four-quark
operators appearing in the operator product expansion of two quark
currents \cite{LSC86}.}
We also find it convenient to leave their color structure as it is,
rather than to remove the generators $T^a$ with help of the identity
$T^a_{\alpha\beta} T^a_{\gamma\delta} = \f{1}{2} \delta_{\alpha\delta}
\delta_{\gamma\beta} - \f{1}{6} \delta_{\alpha\beta}
\delta_{\gamma\delta}$.

	In the remaining part of this paper, we shall concentrate on
finding the Wilson coefficients and the anomalous dimension matrix of
the operators $P_1, ..., P_6$, leaving aside the chromomagnetic moment
operator (\ref{magn.mom}). The latter dimension-five operator does not
require counterterms proportional to dimension-six operators. Thus, it
does not affect renormalization group evolution of their Wilson
coefficients. Consequently, the operators $P_1, ..., P_6$ can be
considered separately, which makes the necessary formulae more
compact. We discuss the chromomagnetic moment operator together with
its photonic counterpart up to three loops in other publications
\cite{CMM97.1,CMM97.4}.

\ \\
{\bf 3. Matching conditions for the Wilson coefficients and the
anomalous dimension matrix}

	We now turn to the values of the coefficients $C_1, ...,
C_6$. They are found by matching the SM and effective theory
amplitudes perturbatively in $\al$. When performing the matching, one
sets the renormalization scale $\mu$ close to $M_W$, in order to
reduce size of $\;\ln(M_W/\mu)$ which could otherwise worsen the
perturbative expansion. In the following, we use the
$\overline{MS}$ scheme with fully anticommuting $\gamma_5$, and
take $\mu = M_W$ as the matching scale.

\begin{figure}[ht] 
\centerline{
\epsfysize = 2.5cm
\epsffile{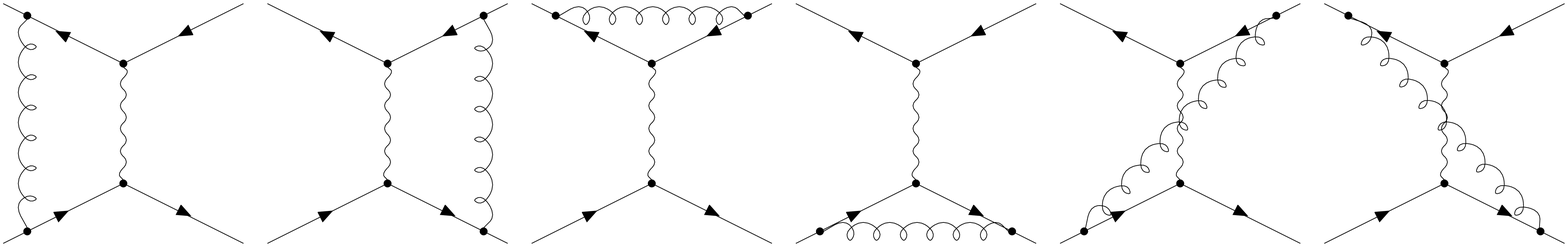}}
\caption{One-loop SM current-current diagrams. 
          \label{1loop.sm.current}}
\end{figure}

\begin{figure}[ht] 
\centerline{
\epsfysize = 2.5cm
\epsffile{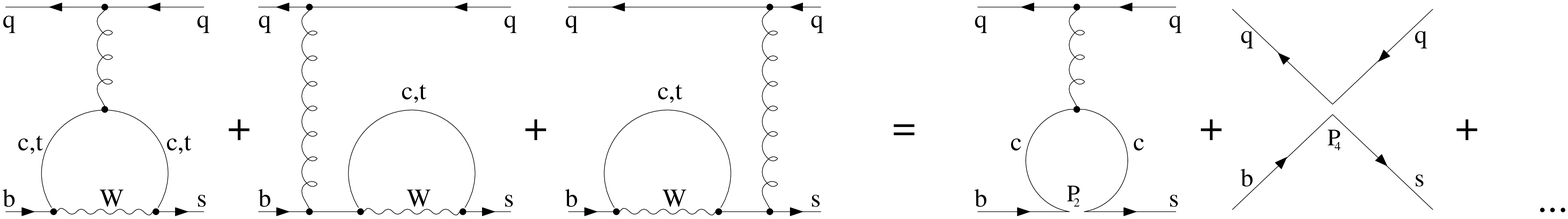}}
\caption{One-loop matching for $C_4$. Dots stand for ${\cal O}(1/M_W^4)$, 
           a counterterm, and for a diagram with insertion of the 
	   chromomagnetic moment operator given in  eqn.~(\ref{magn.mom}).
  \label{penguin.match}}
\end{figure}

	The tree-level matching condition is shown in
fig.~\ref{tree}. The one loop matching conditions for the coefficients
$C_1$ and $C_2$ are found by requiring that the one-loop SM
diagrams in fig.~\ref{1loop.sm.current} be reproduced by the
effective theory diagrams in fig.~\ref{1loop.current}.\footnote{
Tree diagrams with counterterms need to be added on both the SM and
the effective theory sides before the matching is performed.}
The one-loop contribution to the coefficient $C_4$ can be found
by on-shell matching of the 1PR amplitudes presented in
fig.~\ref{penguin.match}. The coefficients $C_3$, $C_5$ and
$C_6$ acquire no one-loop contributions.

	The obtained coefficients $C_i(\mu = M_W)$ read
\bea
C_1(M_W) &=& \f{15 \al}{4 \pi} \;+\; {\cal O}(\al^2), \label{c1m} \\
C_2(M_W) &=& 1 \;+\; {\cal O}(\al^2), \label{c2m} \\
C_4(M_W) &=& \f{\al}{4 \pi} \left[ E(x) - \f{2}{3} \right]
                   \;+\; {\cal O}(\al^2), \label{c4m} \\
C_3(M_W) &=& C_5(M_W) = C_6(M_W) = {\cal O}(\al^2), \label{c356m}
\eea
where
\be
E(x) = \f{x(18-11x-x^2)}{12(1-x)^3} + \f{x^2(15-16x+4x^2)}{6(1-x)^4} \ln x
        -\f{2}{3} \ln x.
\ee
and $x = m_t^2/M_W^2$.

	Many flavor changing processes occur at energy scales
$m_{light} << M_W$. Perturbative expansion of their amplitudes is
efficient when the renormalization scale $\mu$ is close to
$m_{light}$ rather than to $M_W$.  Wilson coefficients $C_i$ at
$\mu \sim m_{light}$ are found from $C_i(M_W)$ with help of the
Renormalization Group Equations (RGE)
\be
\mu \f{d}{d \mu} C_i(\mu) = \sum_{j=1}^6 C_j(\mu) \gamma_{ji}(\mu).
\ee
Here, $\hat{\gamma}(\mu)$ is the anomalous dimension matrix which has
the following perturbative expansion
\be
\hat{\gamma}(\mu) = \f{\al  (\mu)}{ 4 \pi   } \hat{\gamma}^{(0)}
                  \;+\; \f{\al^2(\mu)}{(4 \pi)^2} \hat{\gamma}^{(1)}
		  \;+\; {\cal O}(\al^3).
\ee

	The anomalous dimension matrix in the $MS$ or
$\overline{MS}$ schemes is found from one- and two-loop
counterterms in the effective theory, according to the following
relations:
\bea
\hat{\gamma}^{(0)} &=& 2 \hat{a}^{11}, \label{gamma0} \\
\hat{\gamma}^{(1)} &=& 4 \hat{a}^{12} - 2 \hat{b} \hat{c}.
						\label{gamma1}
\eea
The matrices $\hat{a}^{11}$, $\hat{a}^{12}$ and $\hat{b}$ in the above
equations parameterize the $MS$-scheme counterterms proportional to
$C_1,..., C_6$ in eqn.~(\ref{heff})
\bea \label{counterterms}
(\mbox{counterterms})_i &=& \vspace{0.2cm}
\f{\al}{4 \pi \e} \left[ \sum_{k=1}^6 a^{11}_{ik} P_k 
                   \;+\; \sum_{k=1}^4 b_{ik} E^{(1)}_k 
		   \;+\; n_i N_1 \right] 
\;+\; \f{\al^2}{(4 \pi)^2} \sum_{k=1}^6 
\left( \f{1}{\e^2} a^{22}_{ik} + \f{1}{\e} a^{12}_{ik} \right) P_k \nonumber \\ 
&+& \; (\mbox{two-loop chromomagnetic counterterm}) \nonumber \\ 
&+& \; (\mbox{two-loop nonphysical counterterms}) + {\cal O}(\al^3). 
\eea

\begin{figure}[ht] 
\centerline{
\epsfysize = 3cm
\epsffile{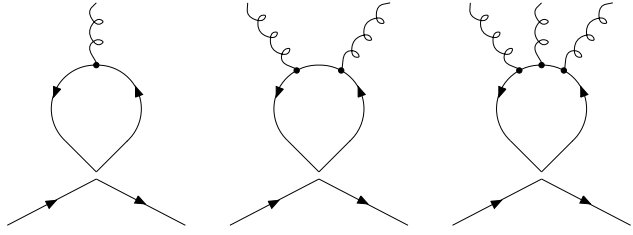}}
\caption{One-loop penguin diagrams with Dirac traces. The traces
		contain no $\gamma_5$. \label{1loop.front.penguin}}
\end{figure}

	The matrix $\hat{c}$ occurs in the one-loop matrix elements of
evanescent operators. Let us denote by $\me{E^{(1)}_k}_{1loop}$ any
one-loop amplitude with an insertion of some evanescent operator
$E^{(1)}_k$. Dependently of what external lines are chosen, such an
amplitude can be given by the diagrams in fig.~\ref{1loop.current} or
in figs.~\ref{1loop.side.penguin} and \ref{1loop.front.penguin}. Pole
parts of such amplitudes are proportional to evanescent operators. The
remaining parts in the limit $D \to 4$ equal to linear combinations of
tree-level matrix elements of the physical operators and $N_1$.
\bea \label{evan.melem}
\me{E^{(1)}_k}_{1loop} &=& \;-\; \sum_{j=0}^4 \f{1}{\e} \left[ 
d_{kj} \me{E_j^{(1)}}_{tree} + e_{kj} \me{E_j^{(2)}}_{tree} \right]
\;-\; \sum_{i=0}^6 c_{ki} \me{P_i}_{tree} \nonumber \\
&& \;+\; x_k \me{(\bar{s}_L \sigma^{\mu\nu} T^a b_R) G^a_{\mu\nu}}_{tree}
   \;-\; \tilde{n}_k \me{N_1}_{tree} \;+\; {\cal O}(\e).
\eea
The coefficients in these linear combinations (which define the matrix
$\hat{c}$) are independent of what particular external lines are
chosen. However, recovering the matrix $\hat{c}$ usually requires
considering several choices of the external lines, so that all the
potentially relevant operators can enter into eqn.~(\ref{evan.melem})
with nonvanishing tree-level matrix elements.

	In ref.~\cite{evan}, one can find a more detailed
explanation why evanescent operators affect the anomalous dimension
matrix precisely in the manner we have just described.

	The matrices $\hat{a}^{11}$, $\hat{a}^{12}$, $\hat{b}$ and
$\hat{c}$ which determine $\hat{\gamma}^{(0)}$ and
$\hat{\gamma}^{(1)}$ are found by evaluating various one- and two-loop
Feynman diagrams with insertions of the operators $P_i$ and
$E^{(1)}_k$. The one-loop diagrams are shown in
figs.~\ref{1loop.current}, \ref{1loop.side.penguin} and
\ref{1loop.front.penguin}. All the two-loop diagrams we calculate
(where only $P_i$ need to be inserted) are identical to 1PI parts of
the two-loop diagrams shown in the figures of ref.~\cite{BJLW93}. Most
of them can be obtained from the one-loop ones by adding an additional
internal gluon.

	We evaluate pole parts of these diagrams using the method we
have described in refs.~\cite{MM95,CMM97.2} where a common mass
parameter was used as an infrared regulator. Next, we decompose the
obtained ultraviolet divergences into terms proportional to the
physical and nonphysical operators. This gives us the necessary
renormalization constants and, in the end, the anomalous dimension
matrix for the operators $P_1,...,P_6$.

	We find (both in the $MS$ and $\overline{MS}$ schemes)
\be \label{gamma0.expl}
\hat{\gamma}^{(0)} = \left[ \begin{array}{cccccc}
\vspace{0.2cm}
-4 & \f{8}{3} &       0     &        -\f{2}{9}         &     0     &      0    \\
\vspace{0.2cm}
12 &     0    &       0     &         \f{4}{3}         &     0     &      0    \\
\vspace{0.2cm}
 0 &     0    &       0     &       -\f{52}{3}         &     0     &      2    \\
\vspace{0.2cm}
 0 &     0    &  -\f{40}{9} &  -\f{160}{9} + \f{4}{3}f &  \f{4}{9} &  \f{5}{6} \\
\vspace{0.2cm}
 0 &     0    &       0     &      -\f{256}{3}         &     0     &     20    \\
\vspace{0.2cm}
 0 &     0    & -\f{256}{9} & -\f{544}{9} + \f{40}{3}f & \f{40}{9} & -\f{2}{3} 
\end{array} \right]
\ee
and 
\be \label{gamma1.expl}
\hat{\gamma}^{(1)} = \left[ \begin{array}{cccccc}
\vspace{0.2cm}
-\f{145}{3}+\f{16}{9}f & -26+\f{40}{27}f &       -\f{1412}{243}         &         -\f{1369}{243}        &         \f{134}{243}         &      -\f{35}{162}            \\ 
\vspace{0.2cm}	    
    -45+\f{20}{3}f     &   -\f{28}{3}    &        -\f{416}{81}          &          \f{1280}{81}         &          \f{56}{81}          &       \f{35}{27}             \\  
\vspace{0.2cm}	    
           0           &        0        &        -\f{4468}{81}         &   -\f{29129}{81}-\f{52}{9}f   &          \f{400}{81}         &   \f{3493}{108}-\f{2}{9}f    \\
\vspace{0.2cm}	    
           0           &        0        & -\f{13678}{243}+\f{368}{81}f & -\f{79409}{243}+\f{1334}{81}f &   \f{509}{486}-\f{8}{81}f    &  \f{13499}{648}-\f{5}{27}f   \\
\vspace{0.2cm}	    
           0           &        0        &  -\f{244480}{81}-\f{160}{9}f &  -\f{29648}{81}-\f{2200}{9}f  &   \f{23116}{81}+\f{16}{9}f   &   \f{3886}{27}+\f{148}{9}f   \\
\vspace{0.2cm}	    
           0           &        0        & \f{77600}{243}-\f{1264}{81}f &  -\f{28808}{243}+\f{164}{81}f & -\f{20324}{243}+\f{400}{81}f & -\f{21211}{162}+\f{622}{27}f
\end{array} \right].
\ee
where $f$ is the number of active flavors (equal to 5 for $\mu \in [m_b,M_W]$).

	In the end of this section, we shall use the obtained
anomalous dimension matrix to find explicit expressions for the Wilson
coefficients 
\be
C_i(\mu) = C^{(0)}_i(\mu) + \f{\al(\mu)}{4 \pi} C^{(1)}_i(\mu) + {\cal O}(\al^2)
\ee
at the renormalization scale $\mu_b \sim m_b$ which is
appropriate for studying b-quark decays. Using the general
solution of the next-to-leading RGE (given eg. in
ref.~\cite{BJLW92}), we find
\bea
C^{(0)}_i(\mu_b) &=& \sum_{j=1}^6 A_{ij} \eta^{a_j}, \\
C^{(1)}_i(\mu_b) &=& \sum_{j=1}^6 
\left[ B_{ij} + B'_{ij}\eta + B^{\scs E}_{ij} E(x) \eta \right] \eta^{a_j}.
\eea
where $\eta = \al(M_W)/\al(\mu_b)$ and
\bea
\vspace{0.4cm}
\vec{a} &\simeq& \left[ \begin{array}{cccccc}
\;\; \f{6}{23} &\;\; -\f{12}{23} &\;\; 0.4086 &\; -0.4230 &\; -0.8994 &\; 0.1456 \;
\end{array} \right],\\
\vspace{0.2cm}
\hat{A} &\simeq& \left[ \begin{array}{cccccc}
\vspace{0.2cm}
     1      &     -1      &     0   &     0   &     0   &     0   \\
\vspace{0.2cm}
 \f{2}{3}   &  \f{1}{3}   &     0   &     0   &     0   &     0   \\
\vspace{0.2cm}
 \f{2}{63}  & -\f{1}{27}  & -0.0659 &  0.0595 & -0.0218 &  0.0335 \\
\vspace{0.2cm}
 \f{1}{21}  &  \f{1}{9}   &  0.0237 & -0.0173 & -0.1336 & -0.0316 \\
\vspace{0.2cm}
-\f{1}{126} &  \f{1}{108} &  0.0094 & -0.0100 &  0.0010 & -0.0017 \\
-\f{1}{84}  & -\f{1}{36}  &  0.0108 &  0.0163 &  0.0103 &  0.0023 
\end{array} \right], \vspace{0.4cm} \\
\hat{B} &\simeq& \left[ \begin{array}{cccccc}
 5.9606 &  1.0951 &  0      &  0      &  0      &  0      \\
 1.9737 & -1.3650 &  0      &  0      &  0      &  0      \\
-0.5409 &  1.6332 &  1.6406 & -1.6702 & -0.2576 & -0.2250 \\
 2.2203 &  2.0265 & -4.1830 & -0.7135 & -1.8215 &  0.7996 \\
 0.0400 & -0.1860 & -0.1669 &  0.1887 &  0.0201 &  0.0304 \\
-0.2614 & -0.1918 &  0.4197 &  0.0295 &  0.1474 & -0.0640 
\end{array} \right], \vspace{0.4cm} \\
\hat{B}' &\simeq& \left[ \begin{array}{cccccc}
 2.0394 &  5.9049 &  0      &  0      &  0      &  0      \\
 1.3596 & -1.9683 &  0      &  0      &  0      &  0      \\
 0.0647 &  0.2187 & -0.2979 & -0.6218 &  0.1880 & -0.1318 \\
 0.0971 & -0.6561 &  0.1071 &  0.1806 &  1.1520 &  0.1242 \\
-0.0162 & -0.0547 &  0.0423 &  0.1041 & -0.0085 &  0.0067 \\
-0.0243 &  0.1640 &  0.0489 & -0.1700 & -0.0889 & -0.0091 
\end{array} \right], \vspace{0.4cm} \\
\hat{B}^{\scs E} &\simeq& \left[ \s \begin{array}{cccccc} 
 0     &\s 0     &\p  0      &  0      &  0      &  0      \\
 0     &\s 0     &\p  0      &  0      &  0      &  0      \\
 0     &\s 0     &\p -0.1933 &  0.1579 &  0.1428 & -0.1074 \\
 0     &\s 0     &\p  0.0695 & -0.0459 &  0.8752 &  0.1012 \\
 0     &\s 0     &\p  0.0274 & -0.0264 & -0.0064 &  0.0055 \\
 0     &\s 0     &\p  0.0317 &  0.0432 & -0.0675 & -0.0074 
\end{array} \right].
\eea

	For $\al(M_Z) = 0.118$ and $\mu = 5$~GeV, the ratio $\eta$ at
NLO is $\eta = \al(M_W)/\al(\mu_b) \simeq 0.1203/0.2117 \simeq 0.568$,
and
\be
\begin{array}{cccccccccccc}
\vec{C}(\mu)&=&                       & ( & 
 -0.480 &  1.023 & -0.0045 & -0.0640 &  0.00043 &  0.00091 
                                      & ) & \\ 
            & & + \f{\al(\mu)}{4 \pi} & ( & 
 12.12  & -0.965 &  0.0230 & -0.779  & -0.0093  & -0.0083  
                                      & ) & + {\cal O}(\al^2)\\
            &=&                       & ( & 
 -0.276 &  1.007 & -0.0041 & -0.0771 &  0.00028 &  0.00077 
                                      & ) & + {\cal O}(\al^2),
\end{array}
\ee
where $m_t = 175$~GeV has been used.

	When processes taking place at energy scales much lower than
$m_b$ are considered, the $b$-quark must be decoupled at $\mu \sim
m_b$. The same refers to the $c$-quark at $\mu \sim m_c$ in the case
of kaon mixing or decays. The number of active flavors $f$ in the
anomalous dimension matrix (\ref{gamma0.expl})--(\ref{gamma1.expl})
changes at the heavy quark thresholds. One-loop matching between
effective theories needs to be performed then (see eg.
ref.~\cite{BLMM94}). Explicit formulae for the Wilson coefficients
become more complex in such cases.

	Having evolved the effective hamiltonian to a low-energy
scale, one calculates its matrix element between some physical
states of interest. The actual way of calculating the matrix
element depends on the process under consideration. One must
always make sure that the same renormalization scheme is used in
the renormalization group evolution and in evaluating the matrix
element.

	We shall not consider any examples of calculating matrix
elements here. Instead, in the next section, we shall demonstrate
that our anomalous dimension matrix is in agreement with the one
found previously in the standard operator basis. Consequently,
phenomenological results obtained in our basis cannot be
different from what is already known. The advantage of
introducing our new basis is not improving the existing two-loop
phenomenology, but rather forming a much more convenient starting
point for even higher loop computations like the one in
refs.~\cite{CMM97.1,CMM97.4}.

\ \\
{\bf 4. Transformation of the anomalous dimension matrix to the
``standard'' basis}

	The present section is devoted to demonstrating that the
anomalous dimension matrix
(\ref{gamma0.expl})--(\ref{gamma1.expl}) found by us in the new
operator basis (\ref{ope}) agrees with the one found previously
in the following ``standard'' basis of operators
\cite{GW79,BJLW92,BJLW93,CFMR94}: 
\be \label{old.ope}
\begin{array}{rl}
O_1 = & (\bar{s}_L^{\alpha} \gamma_{\mu} c_L^{\beta}) 
        (\bar{c}_L^{\beta}  \gamma^{\mu} b_L^{\alpha}),
\vspace{0.2cm} \\
O_2 = & (\bar{s}_L^{\alpha} \gamma_{\mu} c_L^{\alpha}) 
        (\bar{c}_L^{\beta}  \gamma^{\mu} b_L^{\beta}),
\vspace{0.2cm} \\
O_3 = & (\bar{s}_L^{\alpha} \gamma_{\mu} b_L^{\alpha}) 
 \sum_q (\bar{q}_L^{\beta}  \gamma^{\mu} q_L^{\beta}),     
\vspace{0.2cm} \\
O_4 = & (\bar{s}_L^{\alpha} \gamma_{\mu} b_L^{\beta}) 
 \sum_q (\bar{q}_L^{\beta}  \gamma^{\mu} q_L^{\alpha}),     
\vspace{0.2cm} \\
O_5 = & (\bar{s}_L^{\alpha} \gamma_{\mu} b_L^{\alpha}) 
 \sum_q (\bar{q}_R^{\beta}  \gamma^{\mu} q_R^{\beta}),     
\vspace{0.2cm} \\
O_6 = & (\bar{s}_L^{\alpha} \gamma_{\mu} b_L^{\beta}) 
 \sum_q (\bar{q}_R^{\beta}  \gamma^{\mu} q_R^{\alpha}),
\end{array}
\ee
supplemented with the following evanescent operators:
\be \label{old.evan}
\begin{array}{rl}
O^{\scs E}_1 = & 
   (\bar{s}_L^{\alpha} \gamma_{\mu_1}\gamma_{\mu_2}\gamma_{\mu_3} c_L^{\beta}) 
   (\bar{c}_L^{\beta}  \gamma^{\mu_1}\gamma^{\mu_2}\gamma^{\mu_3} b_L^{\alpha})
+(-16+4\e) \; O_1, \vspace{0.2cm} \\
O^{\scs E}_2 = &  
   (\bar{s}_L^{\alpha} \gamma_{\mu_1}\gamma_{\mu_2}\gamma_{\mu_3} c_L^{\alpha}) 
   (\bar{c}_L^{\beta}  \gamma^{\mu_1}\gamma^{\mu_2}\gamma^{\mu_3} b_L^{\beta})
+(-16+4\e) \; O_2, \vspace{0.2cm} \\
O^{\scs E}_3 = & 
   (\bar{s}_L^{\alpha} \gamma_{\mu_1}\gamma_{\mu_2}\gamma_{\mu_3} b_L^{\alpha}) 
\sum_q 
   (\bar{q}_L^{\beta}  \gamma^{\mu_1}\gamma^{\mu_2}\gamma^{\mu_3} q_L^{\beta})
+(-16+4\e) \; O_3, \vspace{0.2cm} \\
O^{\scs E}_4 = & 
   (\bar{s}_L^{\alpha} \gamma_{\mu_1}\gamma_{\mu_2}\gamma_{\mu_3} b_L^{\beta}) 
\sum_q 
   (\bar{q}_L^{\beta}  \gamma^{\mu_1}\gamma^{\mu_2}\gamma^{\mu_3} q_L^{\alpha})
+(-16+4\e) \; O_4, \vspace{0.2cm} \\
O^{\scs E}_5 = & 
   (\bar{s}_L^{\alpha} \gamma_{\mu_1}\gamma_{\mu_2}\gamma_{\mu_3} b_L^{\alpha}) 
\sum_q 
   (\bar{q}_R^{\beta}  \gamma^{\mu_1}\gamma^{\mu_2}\gamma^{\mu_3} q_R^{\beta})
+ (-4-4\e) \; O_5, \vspace{0.2cm} \\
O^{\scs E}_6 = & 
   (\bar{s}_L^{\alpha} \gamma_{\mu_1}\gamma_{\mu_2}\gamma_{\mu_3} b_L^{\beta}) 
\sum_q 
   (\bar{q}_R^{\beta}  \gamma^{\mu_1}\gamma^{\mu_2}\gamma^{\mu_3} q_R^{\alpha})
+ (-4-4\e) \; O_6.
\end{array}
\ee
Such evanescent operators are necessary to renormalize one-loop
current-current diagrams (fig.~\ref{1loop.current}) with
insertions of the operators $O_1,...,O_6$. We have found their
explicit form by imposing the requirements given in eqns.~(4.3) and
(4.8) of ref.~\cite{BJLW93} up to ${\cal O}(\e^2)$.

     We shall pass from the new operator basis
(\ref{ope},\ref{expl.evan.1}) to the old one
(\ref{old.ope},\ref{old.evan}) in a series of subsequent
redefinitions of operators. First, we shall redefine the physical
operators by adding some evanescent ones to them. Next, some of
the evanescent operators will be redefined by adding ``$\e
\times \mbox{(physical operators)}$'' to them. The final (and
easiest) step will be an $\e$-independent linear transformation
of operators.

	Before going into details, let us mention how the
renormalization constants and anomalous dimensions transform when
passing from one operator basis to another. Let us start with a set of
physical operators $\{P_i\}$ and a set of evanescent ones
$\{E_k^{(1)}\}$. The quantities we would like to consider are the one-
and two-loop anomalous dimensions $\hat{\gamma}^{(0)}$ and
$\hat{\gamma}^{(1)}$, the one-loop renormalization constants
$\hat{b}$ and $\hat{d}$, as well as the matrix
$\hat{c}$ (see eqns.~(\ref{counterterms}) and
(\ref{evan.melem})).

	Suppose we want to pass to a new (primed) operator
basis which differs from the initial one by adding some linear
combinations of evanescent operators to the physical ones 
\be \label{first.redef.gen}
P'_i = P_i + \sum_k W_{ik} E^{(1)}_k,
\hspace{4cm}
E'^{(1)}_k = E^{(1)}_k.
\ee
The quantities under consideration transform as follows:
\be \label{gamma.prime}
\hat{\gamma}'^{(0)} = \hat{\gamma}^{(0)},  \hspace{2cm}
\hat{\gamma}'^{(1)} = \hat{\gamma}^{(1)} 
                      - \left[ \hat{W} \hat{c} ,\; \hat{\gamma}^{(0)} \right]
                      - 2 \beta_0 \hat{W} \hat{c},
\ee
\be \label{b.prime}
\hat{b}' = \hat{b} +\hat{W} \hat{d} -\f{1}{2} \hat{\gamma}^{(0)} \hat{W}, 
\hspace{2cm} \hat{c}' = \hat{c},  \hspace{2cm}
\hat{d}' = \hat{d}.
\ee
where $\beta_0 = 11-\f{2}{3}f$ is the well-known one-loop
$\beta$-function coefficient in QCD.

	Next, we redefine the evanescent operators by adding
``$\e \times \mbox{(physical operators)}$'' to them
\be \label{second.redef.gen}
P''_i = P'_i, \hspace{4cm}
E''^{(1)}_k = E'^{(1)}_k + \e \sum_i U_{ki} P'_i.
\ee
The corresponding transformations read
\be \label{gamma.double.prime}
\hat{\gamma}''^{(0)} = \hat{\gamma}'^{(0)}, \hspace{2cm}
\hat{\gamma}''^{(1)} = \hat{\gamma}'^{(1)} 
                   + \left[ \hat{b}' \hat{U},\; \hat{\gamma}'^{(0)} \right]
                   + 2 \beta_0 \hat{b}' \hat{U},
\ee
\be
\hat{b}'' = \hat{b}', \hspace{2cm}
\hat{c}'' = \hat{c}' +\f{1}{2}\hat{U}\hat{\gamma}'^{(0)} -\hat{d}'\hat{U}, 
\hspace{2cm} \hat{d}'' = \hat{d}'.
\ee

	Finally, we perform $\e$-independent linear transformations
of the physical and evanescent operators (separately), passing
to the triply primed basis
\be \label{third.redef.gen}
P'''_i = \sum_j R_{ij} P''_j, \hspace{4cm}
E'''^{(1)}_k = \sum_l M_{kl} E''^{(1)}_l.
\ee
The transformations are now rather trivial
\be \label{gamma.triple.prime}
 \hat{\gamma}'''^{(0)} = \hat{R} \hat{\gamma}''^{(0)} \hat{R}^{-1}, \hspace{2cm}
 \hat{\gamma}'''^{(1)} = \hat{R} \hat{\gamma}''^{(1)} \hat{R}^{-1},
\ee
\be
 \hat{b}''' = \hat{R} \hat{b}'' \hat{M}^{-1}, \hspace{2cm}
 \hat{c}''' = \hat{M} \hat{c}'' \hat{R}^{-1}, \hspace{2cm}
 \hat{d}''' = \hat{M} \hat{d}'' \hat{M}^{-1}, 
\ee

	Before we start to transform our particular basis of the
physical operators $P_i$ (eqn.~(\ref{ope})), and the one-loop
evanescent ones (eqn.~(\ref{expl.evan.1})), we extend it by
introducing four extra evanescent operators 
\be \label{extra.evan}
\begin{array}{rl}
\vspace{0.2cm}
E^{(1)}_5 =& P_5 - 10 P_3 + 6 \tilde{P}_3,\\
\vspace{0.2cm}
E^{(1)}_6 =& P_6 - 10 P_4 + 6 \tilde{P}_4,\\
\vspace{0.2cm}
E^{(1)}_7 =& \tilde{P}_5 - 10 \tilde{P}_3 + 6 P_3,\\
\vspace{0.2cm}
E^{(1)}_8 =& \tilde{P}_6 - 10 \tilde{P}_4 + 6 P_4, \end{array}
\ee
where tilde over an operator denotes inserting $\gamma_5$ in
front of the last quark field. If some operator $P_i$ is a linear
combination of terms $(\bar{s}_L \Gamma b_L)(q \Gamma' q)$, then
$\tilde{P}_i$ is the same linear combination of 
$(\bar{s}_L \Gamma b_L)(q \Gamma' \gamma_5 q)$, where $\Gamma$ 
and $\Gamma'$ stand for arbitrary Dirac matrices.

	The operators $E^{(1)}_5,...,E^{(1)}_8$ are not needed
as counterterms in the initial basis. However, some linear
combinations of them will become parts of either the physical or
the evanescent operators in the triply primed basis which will
be equivalent to the ''standard'' one (eqns.~(\ref{old.ope}) and
(\ref{old.evan})). 

	The anomalous dimension matrices $\hat{\gamma}^{(0)}$
and $\hat{\gamma}^{(1)}$ in the extended basis are just the same
as in eqns.~(\ref{gamma0.expl}) and (\ref{gamma1.expl}). The
matrices $\hat{b}$, $\hat{c}$ and $\hat{d}$ in the extended
basis are found from one-loop diagrams (figs.
\ref{1loop.current}, \ref{1loop.side.penguin} and
\ref{1loop.front.penguin}) with insertions of $P_1,...,P_6$ and
$E^{(1)}_1,...,E^{(1)}_8$. We find
\be \label{b.matrix}
\hat{b} = \left[ \begin{array}{cccccccc}
\vspace{0.2cm}
\f{5}{12} & \f{2}{9} &     0    &     0     & 0 & 0 & 0 & 0\\
\vspace{0.2cm}
     1    &     0    &     0    &     0     & 0 & 0 & 0 & 0\\
\vspace{0.2cm}
     0    &     0    &     0    &     0     & 0 & 0 & 0 & 0\\
\vspace{0.2cm}
     0    &     0    &     0    &     0     & 0 & 0 & 0 & 0\\
\vspace{0.2cm}
     0    &     0    &     0    &     1     & 0 & 0 & 0 & 0\\ 
     0    &     0    & \f{2}{9} & \f{5}{12} & 0 & 0 & 0 & 0
\end{array} \right], \vspace{0.4cm}
\ee
\be \label{c.matrix}
\hat{c} = \left[ \begin{array}{cccccc}
\vspace{0.2cm}
64 & \f{32}{3} &       0      &    \f{4}{9}      &       0      &       0     \\ 
\vspace{0.2cm}
48 &    -64    &       0      &   -\f{8}{3}      &       0      &       0     \\
\vspace{0.2cm}
 0 &      0    & \f{8960}{3}  &     -2432        & -\f{1280}{3} &     320     \\
\vspace{0.2cm}
 0 &      0    & -\f{4480}{9} & -\f{8864}{3}-40f &  \f{640}{9}  & \f{1280}{3} \\
\vspace{0.2cm}
 0 &      0    &  \f{320}{3}  &   -\f{256}{3}    &  -\f{32}{3}  &       8     \\
\vspace{0.2cm}
 0 &      0    & -\f{160}{9}  &  -\f{952}{9}-4f  &  \f{16}{9}   &   \f{32}{3} \\
\vspace{0.2cm}
 ? &      ?    &       ?      &         ?        &       ?      &       ?     \\
 ? &      ?    &       ?      &         ?        &       ?      &       ? 
\end{array} \right], \vspace{0.4cm}
\ee
\be \label{d.matrix}
\hat{d} = \left[ \begin{array}{cccccccc}
\vspace{0.2cm}
-7 & -\f{4}{3} &       0    &     0     & 0 &     0     &     0    &     0    \\
\vspace{0.2cm}
-6 &      0    &       0    &     0     & 0 &     0     &     0    &     0    \\
\vspace{0.2cm}
 0 &      0    & -\f{64}{3} &   -14     & 0 &     0     &     0    &     0    \\
\vspace{0.2cm}
 0 &      0    & -\f{28}{9} & \f{13}{3} & 0 &     0     &     0    &     0    \\
\vspace{0.2cm}
 0 &      0    &       0    &     1     & 0 &     0     &     0    &     6    \\ 
\vspace{0.2cm}
 0 &      0    &   \f{2}{9} & \f{5}{12} & 0 & -\f{9}{2} & \f{4}{3} & \f{5}{2} \\ 
\vspace{0.2cm}
 ? &      ?    &       ?    &     ?     & ? &     ?     &     ?    &     ?    \\ 
 ? &      ?    &       ?    &     ?     & ? &     ?     &     ?    &     ? 
\end{array} \right].
\ee
Question marks have been put into the entries which we do not know,
but which do not affect the quantities calculated below. 

	Transforming two-loop anomalous dimensions to the
''standard'' basis requires knowledge of one-loop diagrams
with insertions of the extra evanescent operators
$E^{(1)}_5,...,E^{(1)}_8$, which introduces traces with
$\gamma_5$ to the calculation. Fortunately, from among all the
relevant quantities listed above, only $c_{64}$ and $c_{84}$
depend on such traces. (In the cases of $c_{54}$ and $c_{74}$,
the relevant color factors vanish.) The trace which occurs in
the case of $c_{64}$ is $Tr(\gamma_{\mu}\gamma_{\nu}\gamma_5)$
which can be safely set to zero. On the other hand, $c_{84}$ is
irrelevant in the particular transformation we perform below.

	At this point, we are ready to start the explicit
transformation. In the first step, we subtract some evanescent
operators from the operators $P_5$ and $P_6$ 
\be \label{first.transf.expl}
P'_5 = P_5 - E^{(1)}_5, \hspace{4cm} P'_6 = P_6 - E^{(1)}_6,
\ee
which is equivalent to taking the $6 \times 8$ matrix $\hat{W}$ in
eqn.~(\ref{first.redef.gen}) equal to
\be \label{W.matrix}
W_{ik} = \left\{ \begin{array}{cc} -1 & \mbox{when } i=k=5
\mbox{ or } i=k=6,\\ 0 & \mbox{otherwise.} \end{array} \right.
\ee

	In the second step, we redefine the evanescent operators
according to eqn.~(\ref{second.redef.gen}), with
\be \label{U.matrix}
\hat{U} = \left[ \begin{array}{cccccc}
\vspace{0.2cm}
4 & 0 &      0     &      0     &     0    &     0    \\
\vspace{0.2cm}
0 & 4 &      0     &      0     &     0    &     0    \\
\vspace{0.2cm}
0 & 0 &      0     &      0     &     0    &     0    \\
\vspace{0.2cm}
0 & 0 &      0     &      0     &     0    &     0    \\
\vspace{0.2cm}
0 & 0 & -\f{20}{3} &      0     & \f{2}{3} &     0    \\ 
\vspace{0.2cm}
0 & 0 &      0     & -\f{20}{3} &     0    & \f{2}{3} \\
\vspace{0.2cm}
0 & 0 &     -4     &      0     &     0    &     0    \\
0 & 0 &      0     &     -4     &     0    &     0 
\end{array} \right].
\ee

	Finally, we perform $\e$-independent linear transformations
according to eqn.~(\ref{third.redef.gen}), with
\be \label{R.matrix}
\hat{R} = \left[ \begin{array}{cccccc}
\vspace{0.2cm}
2 & \f{1}{3} &      0    &      0    &      0     &      0    \\
\vspace{0.2cm}
0 &     1    &      0    &      0    &      0     &      0    \\
\vspace{0.2cm}
0 &     0    & -\f{1}{3} &      0    &  \f{1}{12} &      0    \\
\vspace{0.2cm}
0 &     0    & -\f{1}{9} & -\f{2}{3} &  \f{1}{36} &  \f{1}{6} \\
\vspace{0.2cm}
0 &     0    &  \f{4}{3} &      0    & -\f{1}{12} &      0    \\
0 &     0    &  \f{4}{9} &  \f{8}{3} & -\f{1}{36} & -\f{1}{6}
\end{array} \right]
\ee
and
\be
\hat{M} = \left[ \begin{array}{cccccccc}
\vspace{0.2cm}
2 & \f{1}{3} & 0 & 0 &     0    & 0 &      0    &  0 \\
\vspace{0.2cm}
0 &     1    & 0 & 0 &     0    & 0 &      0    &  0 \\
\vspace{0.2cm}
0 &     0    & 0 & 0 & \f{1}{2} & 0 & -\f{1}{2} &  0 \\
\vspace{0.2cm}
0 &     0    & 0 & 0 & \f{1}{6} & 1 & -\f{1}{6} & -1 \\
\vspace{0.2cm}
0 &     0    & 0 & 0 & \f{1}{2} & 0 &  \f{1}{2} &  0 \\
\vspace{0.2cm}
0 &     0    & 0 & 0 & \f{1}{6} & 1 &  \f{1}{6} &  1 \\
\vspace{0.2cm}
0 &     0    & 1 & 0 &     0    & 0 &      0    &  0 \\
0 &     0    & 0 & 1 &     0    & 0 &      0    &  0 
\end{array} \right].
\ee
It is only a matter of tedious but simple algebra to verify that
the triply primed basis is identical to the ''standard'' one,
i.e. $P'''_i = O_i$ and $E'''^{(1)}_i = O^{\scs E}_i$ for $i
=1,...,6$. The operators $E'''^{(1)}_7$ and $E'''^{(1)}_8$ play role
of extra (unnecessary) evanescent operators in the triply primed
basis, i.e. the physical operators do not mix into them at one
loop.

	Finding the anomalous dimension matrix in the ''standard''
basis is now only a matter of simple matrix multiplication. Combining
eqns.~(\ref{gamma.prime}), (\ref{b.prime}), (\ref{gamma.double.prime})
and (\ref{gamma.triple.prime}), one finds
\bea
\hat{\gamma}'''^{(0)} &=& \hat{R} \hat{\gamma}^{(0)} \hat{R}^{-1},\\
\hat{\gamma}'''^{(1)} &=& \hat{R} \left\{ \hat{\gamma}^{(1)} 
+ \left[ \Delta \hat{r},\; \hat{\gamma}^{(0)} \right]
+ 2 \beta_0 \; \Delta \hat{r} \right\} \hat{R}^{-1}, \label{resemble}
\eea
where 
\be \label{deltar}
\Delta \hat{r} = \left( \hat{b} + \hat{W} \hat{d}
        -\f{1}{2} \hat{\gamma}^{(0)} \hat{W} \right) \hat{U} - \hat{W} \hat{c}.
\ee
The formula (\ref{resemble}) resembles eqn.~(3.4) of
ref.~\cite{BJLW92} where renormalization-scheme dependence of
$\gamma^{(1)}$ was considered in general. 

	Substituting the matrices from eqns.~(\ref{gamma0.expl}),
(\ref{gamma1.expl}), (\ref{b.matrix}), (\ref{c.matrix}),
(\ref{d.matrix}), (\ref{W.matrix}), (\ref{U.matrix}) and
(\ref{R.matrix}) to the above equations, one obtains
\be
\hat{\gamma}'''^{(0)} = \left[ \begin{array}{cccccc}
\vspace{0.2cm}	    
-2 &  6 &      0       &      0        &      0     &     0          \\ 
\vspace{0.2cm}	    
 6 & -2 & -\f{2}{9}    &  \f{2}{3}     & -\f{2}{9}  & \f{2}{3}       \\ 
\vspace{0.2cm}	    
 0 &  0 & -\f{22}{9}   &  \f{22}{3}    & -\f{4}{9}  & \f{4}{3}       \\ 
\vspace{0.2cm}	    
 0 &  0 & 6 -\f{2}{9}f & -2+\f{2}{3}f  & -\f{2}{9}f & \f{2}{3}f      \\ 
\vspace{0.2cm}	    
 0 &  0 &      0       &       0       &      2     &     -6         \\ 
 0 &  0 & -\f{2}{9}f   &  \f{2}{3}f    & -\f{2}{9}f & -16+ \f{2}{3}f 
\end{array} \right] 
\ee
and
\be
\hat{\gamma}'''^{(1)} = \left[ \begin{array}{cccccc}
\vspace{0.2cm}	    
-\f{21}{2}-\f{2}{9}f &  \f{7}{2}+\f{2}{3}f  &          \f{79}{9}            &         -\f{7}{3}          &         -\f{65}{9}         &         -\f{7}{3}            \\ 
\vspace{0.2cm}	    
 \f{7}{2}+\f{2}{3}f  & -\f{21}{2}-\f{2}{9}f &        -\f{202}{243}          &         \f{1354}{81}       &      -\f{1192}{243}        &        \f{904}{81}           \\ 
\vspace{0.2cm}	    
         0           &            0         &   -\f{5911}{486}+\f{71}{9}f   &  \f{5983}{162}+\f{1}{3}f   & -\f{2384}{243}-\f{71}{9}f  &    \f{1808}{81}-\f{1}{3}f    \\ 
\vspace{0.2cm}	    
         0           &            0         &   \f{379}{18}+\f{56}{243}f    &  -\f{91}{6}+\f{808}{81}f   & -\f{130}{9}-\f{502}{243}f  & -\f{14}{3}+\f{646}{81}f      \\ 
\vspace{0.2cm}	    
         0           &            0         &         -\f{61}{9}f           &        -\f{11}{3}f         &    \f{71}{3}+\f{61}{9}f    &     -99+\f{11}{3}f          \\ 
         0           &            0         &        -\f{682}{243}f         &        \f{106}{81}f        & -\f{225}{2}+\f{1676}{243}f & -\f{1343}{6} + \f{1348}{81}f 
\end{array} \right] 
\ee
The above anomalous dimension matrices are in agreement with those found
in refs.~\cite{BJLW92,BJLW93,CFMR94} in the so-called NDR scheme with
fully anticommuting $\gamma_5$ (see eg. eqns.~(4.5) and (4.8) in
ref.~\cite{BJLW92}).

	Matching conditions in the ''standard'' basis can be obtained
from eqns.~(\ref{c1m})--(\ref{c356m}), according to
\be
\vec{C}'''(M_W) = \left( \hat{R}^{-1} \right)^{\scs T} 
	\left[ 1 - \f{\al}{4 \pi} \Delta \hat{r}^{\scs T} \right] \vec{C}(M_W)
	+ {\cal O}(\al^2),
\ee
with $\Delta \hat{r}$ given in eqn.~(\ref{deltar}). The above equation yields
\be
C'''_1(M_W) = \f{11 \al}{8 \pi} + {\cal O}(\al^2),
\hspace{3cm}
C'''_2(M_W) = 1 - \f{11 \al}{24 \pi} + {\cal O}(\al^2),
\ee
\be
C'''_3(M_W) = -\f{1}{3} C'''_4(M_W) = C'''_5(M_W) = -\f{1}{3} C'''_6(M_W) = 
 -\f{\al}{24 \pi} \left( E(x) - \f{2}{3} \right) + {\cal O}(\al^2),
\ee
which agrees with refs.~\cite{BJLW92,BJLW93,CFMR94} as well (see
eg. eqns. (5.5) and (5.6) in ref \cite{BJLW92}).

\ \\
{\bf 5. Summary}

	The $|\Delta F| = 1$ nonleptonic effective hamiltonian has
been considered in a renormalization scheme which allows to
consistently use fully anticommuting $\gamma_5$ at any number of
loops, but at the leading order in the Fermi coupling $G_F$. It is
possible in the specific operator basis we have introduced. Such a
basis is particularly suited for performing multiloop calculations.

	We have evaluated two-loop anomalous dimensions of the
relevant operators and one-loop matching conditions for their Wilson
coefficients in the new basis. Renormalization group evolution of the
coefficients has been studied numerically in the case $\Delta
B=-\Delta S=1$. In the end, the two-loop anomalous dimension
matrix found by us was transformed to the previously used
operator basis, and the existing NLO results have been
confirmed.

\ \\
{\bf Acknowledgments}

	M.~Misiak thanks for hospitality at the University of Zurich
where most of this research has been performed.  K.~Chetyrkin
appreciates the warm hospitality of the Theoretical group of the Max
Planck Institute in Munich where part of this work has been made.

	This work has been partially supported by the German
Bundesministerium f{\"u}r Bildung and Forschung under the contract 06 TM
874 and DFG Project Li 519/2-2. K.~Chetyrkin has been partially
supported by INTAS under Contract INTAS-93-0744-ext. M.~Misiak has been
supported in part by Schweizerischer Nationalfonds, by the Polish
Committee for Scientific Research (under grant 2~P03B~180~09,
1995--97) and by the EC contract HCMP CT92004.
 
\ \\
{\bf Appendix}

	Here, we give the nonphysical operators relevant in
our discussion. In the process of renormalizing off-shell
two-loop amplitudes with insertions of operators $P_1, ... P_6$,
there arise many EOM-vanishing operators, i.e. operators which
vanish by the QCD equations of motion.  However, the specific
structure of only one of them
\be
N_1 = \f{1}{g} (\bar{s}_L T^a \gamma_{\mu} b_L ) D_{\nu} G^{a\;\mu\nu} 
\;\; + \;\; P_4
\ee
is relevant in finding one- and two-loop mixing of the operators given
in eqn.~(\ref{ope}).

\begin{figure}[ht] 
\centerline{
\epsfysize = 2cm
\epsffile{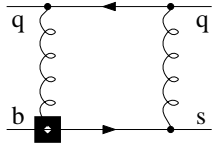}}
\caption{ Sample box diagram with insertion of the first term in $N_1$
            (denoted by the black square). \label{box}}
\end{figure}

	Introduction of $N_1$ in section 2 was the reason why
$P_4$ (and, consequently, $P_3$, $P_5$ and $P_6$) entered the
considerations. However, if $P_4$ was not introduced this way,
$P_3,...,P_6$ would be generated anyway as counterterms to box
diagrams (like the one in fig.~\ref{box}) with insertions of the
first term in $N_1$. This term would be then treated as a
physical operator. Introducing the nonphysical operator $N_1$
makes considering such box diagrams unnecessary, because the
whole $N_1$ cannot have physical counterterms \cite{P80,C84,S94}.

	Another kind of nonphysical operators which we need to
list here are the so-called evanescent operators i.e. operators
which algebraically vanish in 4 dimensions \cite{C84,evan}.

	The one-loop diagrams shown in fig.~\ref{1loop.current} with
insertions of $P_1, ..., P_6$ require counterterms proportional to $P_1,
..., P_6$ themselves and to the following evanescent operators:
\be \label{expl.evan.1}
\begin{array}{rl}
\vspace{0.2cm}
E^{(1)}_1 =& (\bar{s}_L \gamma_{\mu_1}
                      \gamma_{\mu_2}
                      \gamma_{\mu_3} T^a c_L)(\bar{c}_L \gamma^{\mu_1} 
                                                        \gamma^{\mu_2}
                                                        \gamma^{\mu_3} T^a b_L) 
-16 P_1, \\
\vspace{0.2cm}
E^{(1)}_2 =& (\bar{s}_L \gamma_{\mu_1}
                      \gamma_{\mu_2}
                      \gamma_{\mu_3}     c_L)(\bar{c}_L \gamma^{\mu_1} 
                                                        \gamma^{\mu_2}
                                                        \gamma^{\mu_3}     b_L) 
-16 P_2, \\
\vspace{0.2cm}
E^{(1)}_3 =& (\bar{s}_L \gamma_{\mu_1}
                      \gamma_{\mu_2}
                      \gamma_{\mu_3}
                      \gamma_{\mu_4}
                      \gamma_{\mu_5}     b_L)\sum_q(\bar{q} \gamma^{\mu_1} 
                                                            \gamma^{\mu_2}
                                                            \gamma^{\mu_3}
                                                            \gamma^{\mu_4}
                                                            \gamma^{\mu_5}     q) 
-20 P_5 + 64 P_3, \\
\vspace{0.2cm}
E^{(1)}_4 =& (\bar{s}_L \gamma_{\mu_1}
                      \gamma_{\mu_2}
                      \gamma_{\mu_3}
                      \gamma_{\mu_4}
                      \gamma_{\mu_5} T^a b_L)\sum_q(\bar{q} \gamma^{\mu_1} 
                                                            \gamma^{\mu_2}
                                                            \gamma^{\mu_3}
                                                            \gamma^{\mu_4}
                                                            \gamma^{\mu_5} T^a q) 
-20 P_6 + 64 P_4. \end{array}
\ee

	At the two-loop level, we encounter four more evanescent operators
\be \label{expl.evan.2}
\begin{array}{rl}
\vspace{0.2cm}
E^{(2)}_1 =& (\bar{s}_L \gamma_{\mu_1}
                      \gamma_{\mu_2}
                      \gamma_{\mu_3}
                      \gamma_{\mu_4}
                      \gamma_{\mu_5} T^a c_L)(\bar{c}_L \gamma^{\mu_1} 
                                                        \gamma^{\mu_2}
                                                        \gamma^{\mu_3}
                                                        \gamma^{\mu_4}
                                                        \gamma^{\mu_5} T^a b_L) 
-20 E^{(1)}_1 - 256 P_1, \\
\vspace{0.2cm}
E^{(2)}_2 =& (\bar{s}_L \gamma_{\mu_1}
                      \gamma_{\mu_2}
                      \gamma_{\mu_3}
                      \gamma_{\mu_4}
                      \gamma_{\mu_5} T   c_L)(\bar{c}_L \gamma^{\mu_1} 
                                                        \gamma^{\mu_2}
                                                        \gamma^{\mu_3}
                                                        \gamma^{\mu_4}
                                                        \gamma^{\mu_5}     b_L) 
-20 E^{(1)}_2 - 256 P_2, \\
\vspace{0.2cm}
E^{(2)}_3 =& (\bar{s}_L \gamma_{\mu_1}
                      \gamma_{\mu_2}
                      \gamma_{\mu_3}
                      \gamma_{\mu_4}
                      \gamma_{\mu_5}
                      \gamma_{\mu_6}
                      \gamma_{\mu_7}     b_L)\sum_q(\bar{q} \gamma^{\mu_1} 
                                                            \gamma^{\mu_2}
                                                            \gamma^{\mu_3}
                                                            \gamma^{\mu_4}
                                                            \gamma^{\mu_5}
                                                            \gamma^{\mu_6}
                                                            \gamma^{\mu_7}     q) 
-336 P_5 + 1280 P_3, \\
\vspace{0.2cm}
E^{(2)}_4 =& (\bar{s}_L \gamma_{\mu_1}
                      \gamma_{\mu_2}
                      \gamma_{\mu_3}
                      \gamma_{\mu_4}
                      \gamma_{\mu_5}
                      \gamma_{\mu_6}
                      \gamma_{\mu_7} T^a b_L)\sum_q(\bar{q} \gamma^{\mu_1} 
                                                            \gamma^{\mu_2}
                                                            \gamma^{\mu_3}
                                                            \gamma^{\mu_4}
                                                            \gamma^{\mu_5}
                                                            \gamma^{\mu_6}
                                                            \gamma^{\mu_7} T^a q) 
-336 P_6 + 1280 P_4. \end{array}
\ee

	The latter evanescent operators
$E^{(2)}_1,...,E^{(2)}_4$ are necessary in renormalizing
the two-loop amplitudes from which one determines next-to-leading
anomalous dimensions. They are also present in
eqn.~(\ref{evan.melem}) which determines the matrix $\hat{c}$.
However, a lot can be changed in their particular structure
without affecting the two-loop anomalous dimension matrix of the
physical operators. For instance, adding ``$\e \times
(\mbox{physical operator})$'' to $E^{(2)}_k$ affects
$\hat{a}^{12}$ and $\hat{c}$ in eqn.~(\ref{gamma1}), but leaves
$\hat{\gamma}^{(1)}$ unchanged. This is in contrary to what
happens when such a redefinition is applied to the one-loop
evanescent operators $E^{(1)}_k$ (see
eqns.~(\ref{second.redef.gen}) and (\ref{gamma.double.prime})).

	In section 4, the transformation of the two-loop anomalous
dimension matrix of the physical operators has been successfully
performed without specifying what two-loop evanescent operators were
used in the ``standard'' basis. This is another illustration of
insensitivity of two-loop results to the specific structure of
$E^{(2)}_k$. These operators become more important at the three-loop
level.

\newpage
\setlength {\baselineskip}{0.2in}
 
\end{document}